\documentclass[conference,letter]{IEEEtranTCOM}

\usepackage[utf8]{inputenc} 
\usepackage[T1]{fontenc}
\usepackage{url}
\usepackage{ifthen}
\usepackage{cite}
\usepackage{amsmath}%
\usepackage{wasysym}%
\usepackage{amssymb}

\usepackage{mdwmath}
\usepackage{blindtext}
\usepackage{eqparbox}
\DeclareMathOperator{\E}{\mathbb{E}}

\usepackage[english]{babel}
\usepackage{lipsum}
\usepackage{xcolor}
\usepackage{tikz}
\usepackage{graphicx}
\usepackage{algorithm, algorithmic}
\usepackage{siunitx}

\usepackage{calc}
\newcolumntype{M}[1]{>{\centering\arraybackslash}m{#1}}
\newcolumntype{N}{@{}m{0pt}@{}}

\usepackage{geometry}
\DeclareMathAlphabet{\mathbfsl}{OT1}{ppl}{b}{it} 

\newcommand{\be}[1]{\begin{equation}\label{#1}}
\newcommand{\ee}{\end{equation}}

\renewcommand{\leq}{\leqslant}
 
\renewcommand{\geq}{\geqslant}

\newcommand{\Cref}[1]{Co\-ro\-lla\-ry\,\ref{#1}}

\begin{document}
\title{Outage Probability Analysis of Uplink NOMA over Ultra-High-Speed FSO-Backhauled Systems} 

 \author{%
   \IEEEauthorblockN{Mohammad Vahid Jamali{$^1$}, Seyed Mohammad Azimi-Abarghouyi$^2$,~and Hessam Mahdavifar{$^1$}}
   \IEEEauthorblockA{{$^1$}EECS Department, University of Michigan, Ann Arbor, MI 48109, USA (mvjamali@umich.edu, hessam@umich.edu)}
                 \IEEEauthorblockA{{$^2$}Sharif University of Technology, Tehran, Iran  (e-mail: azimi\_sm@ee.sharif.edu)}
}

\maketitle
\begin{abstract}
In this paper, we consider a relay-assisted uplink non-orthogonal multiple access (NOMA) system where two radio frequency (RF) users are grouped for simultaneous transmission, over each resource block, to an intermediate relay which forwards the amplified version of the users' aggregated signals in the presence of multiuser interference to a relatively far destination. In order to cope with the users' ever-increasing desire for higher data rates, a high-throughput free-space optics (FSO) link is employed as the relay-destination backhaul link. Dynamic-order decoding is employed at the destination to determine the priority of the users based on their instantaneous channel state information (CSI). Closed-form expressions for the individual- and sum-rate outage probability formulas are derived in the case of independent Rayleigh fading for the users-relay access links when the FSO backhaul link is subject to Gamma-Gamma turbulence with pointing error.  This work can be regarded as an initial attempt to incorporate power-domain NOMA over ultra-high-speed FSO-backhauled systems, known as mixed RF-FSO systems.
\end{abstract}


\section{Introduction}
Non-orthogonal multiple access (NOMA) is being considered as one of the enabling technologies for the fifth generation (5G) wireless networks. With its two general  power- and code-domain forms, NOMA can potentially pave the way toward higher throughput, lower latency, improved fairness, higher reliability, and massive connectivity \cite{dai2015non}. Motivated by these fascinating advantages, extensive research activities have been carried out in the past few years to promote the NOMA advancement in diverse directions (see, e.g., \cite{ding2017survey} for a comprehensive survey).

In a variety of applications, there is a need to transmit the users' data to a central unit (CU) or a wired base station (BS) while, given the limited power of the users, it is not feasible for the users to directly communicate with the relatively far destination. Motivated by this fact, several recent work have dealt with the relaying problem in downlink and uplink NOMA communications. In particular, capacity analysis of a simple cooperative relaying system, consisting of a source, a relay, and a destination is provided in \cite{kim2015capacity}. The outage probabilities and ergodic sum rate of a downlink two-user NOMA system, with a full-duplex relay helping one of the users, is characterized in \cite{zhong2016non}. Performance of downlink NOMA transmission with an intermediate amplify-and-forward (AF) relay for multiple-antenna systems, and over Nakagami-$m$ fading channels is investigated in \cite{men2015non} and \cite{men2017performance}, respectively. The performance of coordinated direct and relay transmission for two-user downlink and uplink NOMA systems is investigated in \cite{kim2015coordinated} and \cite{kader2018coordinated}, respectively. Hybrid decode-and-forward (DF) and AF relaying in NOMA systems is proposed in \cite{liu2016hybrid}, and forwarding strategy selection is explored in \cite{xiao2018forwarding}.
Additionally, in order to enable NOMA technology for massive communications, primary work on power-domain NOMA can be mixed with the low-complexity recursive approach proposed in \cite{jamali2018low} based on the Kronecker product of NOMA pattern matrices.

All of the aforementioned work consider sub-6 GHz radio frequency (RF) band for the backhaul link through an AF or DF relay in the absence of any external multiuser interference to the NOMA users.
However, the scarce available bandwidth in the sub-6 GHz band will not be able to support the users' aggressive demand for the higher data rates, especially when NOMA is employed in the users-relay access links to provide higher throughput for the users. In this case, the relay-destination backhaul link can pose a severe bottleneck on the end-to-end performance and substantially negate the NOMA advantages through reducing  the users' achievable throughput and reliability which can in turn even increase their latency.

A potential approach to overcome the aforementioned drawback is moving to higher frequency bands, e.g., through the deployment of millimeter-wave \cite{ge20145g} or free-space optics (FSO) backhaul links \cite{demers2011survey}. Millimeter-wave communication is usually preferred for relatively shorter communication lengths due to the severe propagation conditions at millimeter frequencies \cite{andrews2017modeling}. FSO links, on the other hand, can provide much more available bandwidth and support ranges on the order of several kilometers \cite{khalighi2014survey}. 

In this paper, we investigate the outage probability performance of uplink NOMA transmission over mixed RF-FSO systems when an AF relay is employed to forward the amplified received signal from the Rayleigh fading access links to the destination through an ultra-high-throughput directive interference-free FSO link subject to Gamma-Gamma (GG) fading with beam misalignment error. We consider a two-user uplink NOMA system where the communication is subject to the presence of multiuser interference from some independent users. Such  interference can be induced, e.g., due the co-channel interference from some nearby users aiming to communicate with some other relays or destinations. We apply dynamic-order decoding, also employed very recently in \cite{gao2017theoretical,najafi2018non}, to dynamically determine the detection order of the NOMA users at the destination, and then derive the closed-form expressions for the individual- and sum-rate outage probabilities.
This paper can be considered as an initial attempt to incorporate power-domain NOMA in mixed RF-FSO systems.

The rest of the paper is organized as follows. In Section II, we describe the system model. In Section III, we derive the individual- and sum-rate outage probability closed-form formulas for uplink NOMA over mixed-RF-FSO system. Section IV provides the numerical results, and Section V concludes the paper. 
\section{System Model}
Consider two RF users $\mathcal{U}_1$ and  $\mathcal{U}_2$ grouped together for uplink NOMA transmission to an AF relay $\mathcal{R}$. Denote the composite channel gain of the $\mathcal{U}_i-\mathcal{R}$ link by $h_i=\sqrt{L_i}\tilde{h}_i$, $i=1,2$, where $L_i$ and $\tilde{h}_i$ are respectively the path loss gain and independent-and-identically-distributed (iid) Rayleigh fading coefficient of the $\mathcal{U}_i-\mathcal{R}$ RF link given by \cite[Eq. (2)]{jamali2016link}. Furthermore, assume that the uplink transmission to the relay is affected by undesired multiuser interference from $K$ interfering users $\mathcal{I}_k$, $k=1,2,...,K$, each with the transmit power $p'_k$, path loss gain $L'_k$, and iid Rayleigh fading coefficient $\tilde{h}'_k$. This interference can be from the users scheduled for the concurrent transmission to some other relays in the cellular network or any other non-vanishing interference during the desired transmission block. 
The  received signal  by the relay is then expressed as
\begin{align}\label{y_R}
y_{\mathcal{R}}=\sum_{i=1}^{2}x_i\tilde{h}_i\sqrt{a_iL_iP}+\sum_{k=1}^{K}x'_k\tilde{h}'_k\sqrt{L'_kp'_k}+n_{\mathcal{R}},
\end{align}
where $x_i$ and $x'_k$ are the transmit symbols by $\mathcal{U}_i$ and $\mathcal{I}_k$, respectively, and $n_{\mathcal{R}}$ is the additive white Gaussian noise (AWGN) of the relay receiver with mean zero and variance $\sigma^2_{\mathcal{R}}$. Moreover, $a_1$ and $a_2=1-a_1$ are the power allocation coefficients determining the portion of the total power $P$ assigned to each of the desired users. Note that for the users with independent Rayleigh fading, all fading gains $|\tilde{h}_i|^2$'s and $|\tilde{h}'_k|^2$'s have an exponential distribution with mean one (to ensure that fading neither amplifies nor attenuates the received power) as $f_{|\tilde{h}_i|^2}(x)=f_{|\tilde{h}'_k|^2}(x)=\exp(-x)$, $x\geq 0$. Note that, in this paper, for the sake of simplicity, we do not consider the stochastic geographical positions of the nodes, and deal with the building part of the analysis, corresponding to a given time slot, where the nodes are at some fixed positions.

The received signal $y_{\mathcal{R}}$ at the relay is then converted to optical signal using intensity-modulation direct-detection (IM/DD), and is amplified with a constant gain $G$ to keep the disparity between the power level of different NOMA users for successive interference cancellation (SIC) detection at the destination. In this case, the transmitted optical signal by the relay toward the destination $\mathcal{D}$ can be expressed as $S_{\mathcal{R}}=G(1+\eta y_{\mathcal{R}})$, where $\eta$ is the electrical-to-optical conversion coefficient \cite{lee2011performance}.
The transmitted signal then undergoes the composite FSO channel gain of $g=g_l\tilde{g}$ where $g_l$ is the path loss gain of the $\mathcal{R-D}$ FSO backhaul link, with length $d_{\mathcal{RD}}$, defined as $g_l=\rho\times10^{-\kappa d_{\mathcal{RD}}/10}$
where $\rho$ is the responsivity of the photodetector, and $\kappa$ is the weather-dependent attenuation coefficient \cite{jamali2016link}.
Moreover, $\tilde{g}=g_pg_f$ is the total fading coefficient due to pointing error $g_p$ and optical turbulence $g_f$. In the case of GG optical turbulence with beam misalignment, the distribution of $\tilde{g}$ can be expressed as \cite{sandalidis2009optical}
\begin{align}\label{f_gtild}
\! f_{\tilde{g}}(\tilde{g})\!=\!\frac{\alpha\beta\xi^2}{A_0\Gamma(\alpha)\Gamma(\beta)}{\rm G}_{1,3}^{3,0}\left[\!\frac{\alpha\beta}{A_0}\tilde{g}{\bigg |\begin{matrix}
	\xi^2	\\ \xi^2-1,\alpha\!-\!1,\beta\!-\!1
	\end{matrix}}\!\right]\!,\!
\end{align}
where $\alpha$ and $\beta$ are the fading parameters of the GG distribution, $\xi$ is the ratio of the equivalent beam radius and the pointing error
displacement standard deviation (jitter) measured at the receiver, $\Gamma(\cdot)$ is the Gamma function \cite[Eq. (8.310)]{gradshteyn2014table}, and ${\rm G}[\cdot]$ is the Meijer's G-function \cite[Eq. (9.301)]{gradshteyn2014table}. Furthermore, $A_0$ is the geometric loss in the case of perfect beam alignment (zero radial displacement) defined as $A_0=[{\rm erf}(\sqrt{\pi}r/(\sqrt{2}\phi d_{\mathcal{RD}}))]^2$ in which ${\rm erf}(\cdot)$ is the error function, $r$ is the receiver aperture radius, and $\phi$ is the transmitter beam divergence angle. 
The destination then filters out the direct current (DC) component of $g_l\tilde{g}G$ from $g_l\tilde{g}S_{\mathcal{R}}+n_{\mathcal{D}}$ to obtain the received signal as
\begin{align}\label{y_D}
y_{\mathcal{D}}\!=\!\eta g_l\tilde{g}G\!\left(\!\sum_{i=1}^{2}\!x_i\tilde{h}_i\sqrt{\!a_iL_iP}\!+\!\!\sum_{k=1}^{K}\!x'_k\tilde{h}'_k\sqrt{\!L'_kp'_k}\!+\!n_{\mathcal{R}}\!\!\right)\!\!+\!n_{\mathcal{D}},\!
\end{align}
where $n_{\mathcal{D}}$ is the destination AWGN with mean zero and variance $\sigma^2_{\mathcal{D}}$.

We assume that the NOMA users are indexed based on their path loss gains, i.e., $L_1\geq L_2$, and the power allocation strategy proposed in \cite{zhang2016uplink} is adopted to determine $a_1$ and $a_2$ as $a_1L_1=a_2L_2\times10^{s/10}$ where $s\geq 0$ is the power back-off step; hence, $a_1=L_2\times10^{s/10}/(L_1+L_2\times10^{s/10})$, and $a_2=L_1/(L_1+L_2\times10^{s/10})$. We further assume that the BS has perfect knowledge about the channel state information (CSI) and orders the NOMA users based on their instantaneous received power. In fact, based on the principles of uplink power-domain NOMA \cite{zhang2016uplink,yang2016general}, the BS orders the users based on their channel conditions from best to worst. Such a dynamically ordering the users can potentially prevent the possibility of firstly decoding the users with worse instantaneous channel conditions if they are fixedly ordered based on their channel statistics. 
Therefore, depending on the fading coefficients $\tilde{h}_1$ and $\tilde{h}_2$, the detection order is either $\pi_1=(1,2)$, meaning that the first user is decoded first, if $a_1L_1|\tilde{h}_1|^2\geq a_2L_2|\tilde{h}_2|^2$ or $\pi_2=(2,1)$ if otherwise.



\section{Outage Probability Analysis}
In this section, we characterize the individual- and sum-rate outage probability formulas for dual-hop uplink NOMA over mixed RF-FSO systems. 
\subsection{Individual-Rate Outage Analysis}
\begin{figure*}[!t]
	\normalsize
	\vspace{-0.25cm}
	\setcounter{equation}{6}
	\begin{align}\label{Joint1,1}
	\Pr(\gamma^{(1)}_{\pi_1}<\gamma^{(1)}_{\rm th},\pi_1){\Big|}_{\gamma^{(1)}_{\rm th}<1}
	&	\stackrel{(a)}{=}\Pr\left(|\tilde{h}_1|^2<\gamma^{(1)}_{\rm th}\left[|\tilde{h}_2|^2\times10^{-s/10}+{\boldsymbol{\mathcal{I}}_1}+C_{\mathcal{D}}/(a_1L_1P\tilde{g}^2)\right],|\tilde{h}_1|^2\geq |\tilde{h}_2|^2\times10^{-s/10}\right)\nonumber\\
	&\hspace{-3.8cm}\stackrel{(b)}{=}\E_{|\tilde{h}_2|^2\!<\!{J^{(1)}_{\rm th}}\big(\boldsymbol{\mathcal{I}}_1\!+\!C_{\mathcal{D}}/(a_1L_1P\tilde{g}^2)\big)}\!\!\left[\exp\left(-|\tilde{h}_2|^2\!\times\!10^{-s/10}\right)-\exp\left(-\gamma^{(1)}_{\rm th}\!\Big[|\tilde{h}_2|^2\!\times\!10^{-s/10}+{\boldsymbol{\mathcal{I}}_1}+C_{\mathcal{D}}/(a_1L_1P\tilde{g}^2)\Big]\right)\right]\nonumber\\
	=&\E_{\boldsymbol{\mathcal{I}}_1,\tilde{g}}\left[({1+10^{-s/10}})^{-1}\times\left[1-\exp\left(-\left[{1+10^{-s/10}}\right]\!{J^{(1)}_{\rm th}}\big(\boldsymbol{\mathcal{I}}_1\!+\!C_{\mathcal{D}}/(a_1L_1P\tilde{g}^2)\big)\right)\right]\right]\nonumber\\
	&\hspace{-4.3cm} -\!\E_{\boldsymbol{\mathcal{I}}_1,\tilde{g}}\!\!\bigg[\!({1\!+\!\gamma^{(1)}_{\rm th}\!\!\times\!10^{-s/10}})^{-1}\!\times\!\!\left[\!1\!-\!\exp\!\left(\!-\!\left[{1\!+\!\gamma^{(1)}_{\rm th}\!\!\times\!\!10^{-s/10}}\right]\!{J^{(1)}_{\rm th}}\big(\boldsymbol{\mathcal{I}}_1\!+\!C_{\mathcal{D}}/(a_1\!L_1\!P\tilde{g}^2)\big)\right)\right]\!\exp\!\left(\!-\gamma^{(1)}_{\rm th}\!\Big[{\boldsymbol{\mathcal{I}}_1}\!+\!C_{\mathcal{D}}/(a_1\!L_1\!P\tilde{g}^2)\Big]\!\right)\!\!\bigg]\nonumber\\
	&\hspace{-4.1cm}\stackrel{(c)}{=}\frac{10^{s/10}}{1+10^{s/10}}\left(1\!-\!\exp\!\bigg(\frac{-J^{(1)}_{{\rm th},1}\sigma^2_{\mathcal{R}}}{a_1L_1P}\bigg)\E_{\tilde{g}}\!\!\left[\exp\!\bigg(\frac{-J^{(1)}_{{\rm th},1}C_{\mathcal{D}}}{a_1L_1P\tilde{g}^2}\bigg)\right]\prod_{k=1}^{K}\frac{a_1L_1P}{a_1L_1P+{J^{(1)}_{{\rm th},1}}L'_kp'_k}\right)\!-\!\frac{10^{s/10}}{\gamma^{(1)}_{\rm th}+10^{s/10}}\Bigg(\!\!\exp\!\bigg(\frac{-\gamma^{(1)}_{\rm th}\sigma^2_{\mathcal{R}}}{a_1L_1P}\bigg)\nonumber\\
	&\hspace{-4.1cm}\times\E_{\tilde{g}}\!\!\left[\exp\!\bigg(\frac{-\gamma^{(1)}_{\rm th}C_{\mathcal{D}}}{a_1L_1P\tilde{g}^2}\bigg)\right]\prod_{k=1}^{K}\frac{a_1L_1P}{a_1L_1P+{\gamma^{(1)}_{\rm th}}L'_kp'_k}-\exp\bigg(\frac{-\sigma^2_{\mathcal{R}}J^{(1)}_{{\rm th},2}}{a_1L_1P}\bigg)\E_{\tilde{g}}\!\!\left[\exp\!\bigg(\frac{-C_{\mathcal{D}}J^{(1)}_{{\rm th},2}}{a_1L_1P\tilde{g}^2}\bigg)\right]\prod_{k=1}^{K}\frac{a_1L_1P}{a_1L_1P+L'_kp'_kJ^{(1)}_{{\rm th},2}}\Bigg).
	\end{align}
	\hrulefill
	\vspace{-0.25cm}
\end{figure*}
Note that if the detection order is $\pi_1$, the SIC receiver first treats the signal from the second NOMA user as noise to decode $x_1$ with the signal-to-interference-plus-noise ratio (SINR) of 
\setcounter{equation}{3}
\begin{align}\label{gamma_f}
\!\!\gamma^{(1)}_{\pi_1}\!=\!\frac{a_1L_1P\tilde{g}^2|\tilde{h}_1|^2}{a_2L_2P\tilde{g}^2|\tilde{h}_2|^2\!+\!\sum_{k=1}^{K}{L'_kp'_k}\tilde{g}^2|\tilde{h}'_k|^2\!+\!\tilde{g}^2\sigma^2_{\mathcal{R}}\!+\!C_{\mathcal{D}}},
\end{align}
and then, after removing the received power from the first user, decodes $x_2$ with the SINR of
\begin{align}\label{gamma_s}
\gamma^{(2)}_{\pi_1}=\frac{a_2L_2P\tilde{g}^2|\tilde{h}_2|^2}{\sum_{k=1}^{K}{L'_kp'_k}\tilde{g}^2|\tilde{h}'_k|^2+\tilde{g}^2\sigma^2_{\mathcal{R}}+C_{\mathcal{D}}},
\end{align}
where $C_{\mathcal{D}}=\sigma^2_{\mathcal{D}}/(\eta^2 g_l^2G^2)$. Similarly, when the detection order is $\pi_2$  the SINR values $\gamma^{(1)}_{\pi_2}$ and $\gamma^{(2)}_{\pi_2}$ can be obtained by appropriate change of indexes in \eqref{gamma_f} and \eqref{gamma_s}.

The outage probability of the first user $\mathcal{U}_1$ in achieving an individual rate of $R^{(1)}_{\rm th}$ can be characterized as
\begin{align}\label{p_out_1}
\!\!P_{\rm out}^{(1)}&\stackrel{(a)}{=}P(\pi_1)P_{{\rm out}|\pi_1}^{(1)}+P(\pi_2)P_{{\rm out}|\pi_2}^{(1)}\nonumber\\
&\stackrel{(b)}{=} 1\!-\!\Big[\Pr(\gamma^{(1)}_{\pi_1}>\gamma^{(1)}_{\rm th},\pi_1)+\nonumber\\
&\hspace{-0.3cm}\Pr(\gamma^{(2)}_{\pi_2}>\gamma^{(2)}_{\rm th},\pi_2)\times\Pr(\gamma^{(1)}_{\pi_2}>\gamma^{(1)}_{\rm th},\pi_2)/P(\pi_2)\Big],\!
\end{align}
where step $(a)$ follows from the law of total probability by defining  $P_{{\rm out}|\pi_i}^{(1)}$, $i=1,2$, as the conditional outage probability of the first NOMA user given the decoding order $\pi_i$. Moreover, in step $(a)$, $P(\pi_1)=\Pr(|\tilde{h}_1|^2\geq |\tilde{h}_2|^2\times10^{-s/10})=\E_{|\tilde{h}_2|^2}[\exp(-|\tilde{h}_2|^2\times10^{-s/10})]=(1+10^{-s/10})^{-1}$, and $P(\pi_2)=1-P(\pi_1)=(1+10^{s/10})^{-1}$ are the probabilities of having decoding orders $\pi_1$, and $\pi_2$, respectively. Furthermore, step $(b)$ follows, first, by defining $P_{{\rm cov}|\pi_i}^{(1)}=1-P_{{\rm out}|\pi_i}^{(1)}$, $i=1,2$, as the probability of successfully achieving $R^{(1)}_{\rm th}$ for $\mathcal{U}_1$  conditioned on the decoding order $\pi_i$, and then noting that correct detection of $x_1$ for the decoding order $\pi_2=(2,1)$ requires successful decoding of the preceding symbol $x_2$, i.e., $P_{{\rm cov}|\pi_1}^{(1)}=\Pr(\gamma^{(1)}_{\pi_1}>\gamma^{(1)}_{\rm th}|\pi_1)$ and $P_{{\rm cov}|\pi_2}^{(1)}=\Pr(\gamma^{(2)}_{\pi_2}>\gamma^{(2)}_{\rm th}|\pi_2)\times\Pr(\gamma^{(1)}_{\pi_2}>\gamma^{(1)}_{\rm th}|\pi_2)$ where
 $\gamma^{(i)}_{\rm th}$ is the threshold SINR for an IM/DD FSO link to achieve the desired data rate $R^{(i)}_{\rm th}$, $i=1,2$. In the following, we calculate the three joint probabilities in \eqref{p_out_1} to ascertain the outage probability of the first user $\mathcal{U}_1$.

In order to calculate $\Pr(\gamma^{(1)}_{\pi_1}>\gamma^{(1)}_{\rm th},\pi_1)$ we first note that $\Pr(\gamma^{(1)}_{\pi_1}>\gamma^{(1)}_{\rm th},\pi_1)=P(\pi_1)\Pr(\gamma^{(1)}_{\pi_1}>\gamma^{(1)}_{\rm th}|\pi_1)=P(\pi_1)[1-\Pr(\gamma^{(1)}_{\pi_1}<\gamma^{(1)}_{\rm th}|\pi_1)]=P(\pi_1)-\Pr(\gamma^{(1)}_{\pi_1}<\gamma^{(1)}_{\rm th},\pi_1)$.
Then using \eqref{gamma_f}, $\Pr(\gamma^{(1)}_{\pi_1}<\gamma^{(1)}_{\rm th},\pi_1)$ can be calculated as \eqref{Joint1,1} shown at the top of this page where, in step $(a)$, ${\boldsymbol{\mathcal{I}}_1}=(\sum_{k=1}^{K}{L'_kp'_k}|\tilde{h}'_k|^2+\sigma^2_{\mathcal{R}})/(a_1L_1P)$ is the sum of the power of multiuser interference and noise, at the relay, normalized to the average power of the first NOMA user. Moreover, step $(b)$ follows, first, by defining the constant $J^{(1)}_{\rm th}=10^{s/10}\times\gamma^{(1)}_{\rm th}/(1-\gamma^{(1)}_{\rm th})>0$ for $\gamma^{(1)}_{\rm th}<1$, and then noting that $\Pr(X<Y,X\geq Z)$ for three random variables (RVs) $X$, $Y$, and $Z$ can be calculated using the law of total probability as $\Pr(X<Y,X\geq Z)=\Pr(Z\leq X<Y, Z<Y)$ since $\Pr(Z\leq X<Y, Z\geq Y)=0$. Finally, step $(c)$ follows by noting that for any constant $C$
\begin{align}\label{E_1,1}
\setcounter{equation}{7}
&\E_{\boldsymbol{\mathcal{I}}_1,\tilde{g}}\!\left[\Pr\!\left(|\tilde{h}_2|^2\!>\!C\big[\boldsymbol{\mathcal{I}}_1+C_{\mathcal{D}}/(a_1L_1P\tilde{g}^2)\big]{\Big|{\boldsymbol{\mathcal{I}}_1,\tilde{g}^2}}\right)\!\right]
\nonumber\\
&\stackrel{(a)}{=}\E_{\boldsymbol{\mathcal{I}}_1,\tilde{g}}\left[\exp\left(-C\big[\boldsymbol{\mathcal{I}}_1+C_{\mathcal{D}}/(a_1L_1P\tilde{g}^2)\big]\right)\right]\nonumber\\
&\stackrel{(b)}{=}\!\exp\!\bigg(\!\!\frac{-C\sigma^2_{\mathcal{R}}}{a_1L_1P}\!\!\bigg)\!\E_{\tilde{g}}\!\!\left[\!\exp\!\bigg(\!\!\frac{-CC_{\mathcal{D}}}{a_1\!L_1\!P\tilde{g}^2}\!\!\bigg)\!\!\right]\!\!\prod_{k=1}^{K}\!\frac{a_1L_1P}{a_1\!L_1\!P\!+\!C\!L'_kp'_k}\!,\!
\end{align}
where step $(a)$ follows from the cumulative density function (CDF) of exponential distribution $\Pr(|\tilde{h}_2|^2<x)=1-\exp(-x)$, and $(b)$ is obtained using the independency of ${\boldsymbol{\mathcal{I}}}_1$ and $\tilde{g}$, and then applying the independency among $|\tilde{h}'_k|^2$'s to get $\E_{\boldsymbol{\mathcal{I}}_1}\!\big[\exp(-C\boldsymbol{\mathcal{I}}_1)\big]=\exp\big(\frac{-C\sigma^2_{\mathcal{R}}}{a_1L_1P}\big)\prod_{k=1}^{K}\E_{|{\tilde{h}'}_{k}|^2}\left[\exp\left(-CL'_kp'_k|\tilde{h}'_k|^2/(a_1L_1P)\right)\right]$. Furthermore, in step $(c)$, $J^{(1)}_{{\rm th},1}=J^{(1)}_{\rm th}({1+10^{-s/10}})$ and $J^{(1)}_{{\rm th},2}=\gamma^{(1)}_{\rm th}+J^{(1)}_{\rm th}({1+\gamma^{(1)}_{\rm th}\times10^{-s/10}})$.

We should further emphasize that \eqref{Joint1,1} is obtained for $\gamma^{(1)}_{\rm th}<1$. If $\gamma^{(1)}_{\rm th}\geq1$, the upper limit of $|\tilde{h}_1|^2$ in \eqref{Joint1,1} is always greater than its lower limit meaning that the condition $(1-\gamma^{(1)}_{\rm th})|\tilde{h}_2|^2\times10^{-s/10}<\gamma^{(1)}_{\rm th}\left[{\boldsymbol{\mathcal{I}}_1}+C_{\mathcal{D}}/(a_1L_1P\tilde{g}^2)\right]$ holds for the all values of $|\tilde{h}_2|^2$ and there is no need to impose such an extra condition on the calculation of the corresponding probability. As a consequence, $\Pr(\gamma^{(1)}_{\pi_1}<\gamma^{(1)}_{\rm th},\pi_1)$ for $\gamma^{(1)}_{\rm th}\geq1$ can be calculated as \eqref{E1,3} shown at the top of the next page. 

\begin{figure*}[!t]
	\normalsize
	\vspace{-0.25cm}
	\begin{align}\label{E1,3}
	\Pr(\gamma^{(1)}_{\pi_1}<\gamma^{(1)}_{\rm th},\pi_1){\Big |}_{\gamma^{(1)}_{\rm th}\geq1}\!\!{=}\frac{10^{s/10}}{1\!+\!10^{s/10}}-\frac{10^{s/10}}{\gamma^{(1)}_{\rm th}\!+\!10^{s/10}}\exp\!\bigg(\!\frac{-\gamma^{(1)}_{\rm th}\sigma^2_{\mathcal{R}}}{a_1L_1P}\!\bigg)\!\E_{\tilde{g}}\!\!\left[\exp\!\bigg(\frac{-\gamma^{(1)}_{\rm th}C_{\mathcal{D}}}{a_1L_1P\tilde{g}^2}\bigg)\right]\prod_{k=1}^{K}\frac{a_1L_1P}{a_1L_1P\!+\!{\gamma^{(1)}_{\rm th}}L'_kp'_k}.
	\end{align}
	\hrulefill
	\vspace{-0.25cm}
\end{figure*}
\begin{figure*}[!t]
	\normalsize
	\vspace{-0.25cm}
	\begin{align}\label{E_g}
	\mathcal{G}(A)\!=\!\E_{\tilde{g}}\!\left[\exp\!\left(\!-A/\tilde{g}^2\right)\!\right]\!\!=\!\frac{\xi^2\!\times\!2^{\alpha+\beta-2}}{2\pi \Gamma(\alpha)\Gamma(\beta)}{\rm G}_{7,2}^{0,7}\!\left[\!\frac{16A_0^2}{A(\alpha\beta)^2}{\bigg |\begin{matrix}
		1,\!(1\!-\!\xi^2)\!/2,\!(2\!-\!\xi^2)\!/2,\!(1\!-\!\alpha)\!/2,\!(2\!-\!\alpha)\!/2,\!(1\!-\!\beta)\!/2,\!(2\!-\!\beta)\!/2\\ -\xi^2/2,(1-\xi^2)/2
		\end{matrix}}\right]\!.\!
	\end{align}
	\vspace{-0.25cm}
	\hrulefill
\end{figure*}
\begin{figure*}[!t]
	\normalsize
	\vspace{-0.25cm}
	\begin{align}\label{E2,1}
	\Pr(\gamma^{(2)}_{\pi_2}<\gamma^{(2)}_{\rm th},\pi_2){\Big|}_{\gamma^{(2)}_{\rm th}<1}\!&=\frac{1}{1\!+\!10^{s/10}}\Bigg(\!1\!-\!\exp\!\bigg(\frac{-J^{(2)}_{{\rm th},1}\sigma^2_{\mathcal{R}}}{a_2L_2P}\bigg)\mathcal{G}\!\bigg(\frac{J^{(2)}_{{\rm th},1}C_{\mathcal{D}}}{a_2L_2P}\bigg)\!\prod_{k=1}^{K}\frac{a_2L_2P}{a_2L_2P\!+\!{J^{(2)}_{{\rm th},1}}L'_kp'_k}\!\Bigg)-\frac{10^{-s/10}}{\gamma^{(2)}_{\rm th}+10^{-s/10}}\nonumber\\
	&\hspace{-4cm}\times\Bigg[\exp\!\bigg(\frac{-\gamma^{(2)}_{\rm th}\sigma^2_{\mathcal{R}}}{a_2L_2P}\bigg)\mathcal{G}\!\bigg(\frac{\gamma^{(2)}_{\rm th}C_{\mathcal{D}}}{a_2L_2P}\bigg)\prod_{k=1}^{K}\frac{a_2L_2P}{a_2L_2P+{\gamma^{(2)}_{\rm th}}L'_kp'_k}-\exp\!\bigg(\frac{-\sigma^2_{\mathcal{R}}J^{(2)}_{{\rm th},2}}{a_2L_2P}\bigg)\mathcal{G}\!\bigg(\frac{C_{\mathcal{D}}J^{(2)}_{{\rm th},2}}{a_2L_2P}\bigg)\prod_{k=1}^{K}\frac{a_2L_2P}{a_2L_2P+L'_kp'_kJ^{(2)}_{{\rm th},2}}\Bigg].
	\end{align}
	\hrulefill
	\vspace{-0.25cm}
\end{figure*}
\begin{figure*}[!t]
	\normalsize
	\vspace{-0.25cm}
	\begin{align}\label{E2,2}
	\Pr(\gamma^{(2)}_{\pi_2}<\gamma^{(2)}_{\rm th},\pi_2){\Big |}_{\gamma^{(2)}_{\rm th}\geq1}\!\!{=}\frac{1}{1\!+\!10^{s/10}}-\frac{10^{-s/10}}{\gamma^{(2)}_{\rm th}\!+\!10^{-s/10}}\exp\!\bigg(\!\frac{-\gamma^{(2)}_{\rm th}\sigma^2_{\mathcal{R}}}{a_2L_2P}\!\bigg)\mathcal{G}\!\bigg(\frac{\gamma^{(2)}_{\rm th}C_{\mathcal{D}}}{a_2L_2P}\bigg)\prod_{k=1}^{K}\frac{a_2L_2P}{a_2L_2P\!+\!{\gamma^{(2)}_{\rm th}}L'_kp'_k}.
	\end{align}
	\hrulefill
	\vspace{-0.25cm}
\end{figure*}

Finally, using \eqref{Joint1,1} for $\gamma^{(1)}_{\rm th}<1$ or \eqref{E1,3} for $\gamma^{(1)}_{\rm th}\geq1$, one can obtain the coverage probability of the first NOMA user given the decoding order $\pi_1$ as $P_{\rm cov}^{(1)}(\pi_1)=(1+10^{-s/10})^{-1}-\Pr(\gamma^{(1)}_{\pi_1}<\gamma^{(1)}_{\rm th},\pi_1)$. However, the closed-form characterization of $P_{\rm cov}^{(1)}(\pi_1)$ still requires the calculation of expressions of the form $\E_{\tilde{g}}\!\!\left[\exp\left(-A/\tilde{g}^2\right)\right]$, where $A$ is a constant and $\tilde{g}$ is distributed according to \eqref{f_gtild}. To do so, we first apply \cite[Eq. (11)]{adamchik1990algorithm} and \cite[Eq. (9.31.2)]{gradshteyn2014table} to write $\exp\left(-A/\tilde{g}^2\right)$ in the form of Meijer's G-function as $\exp\left(-A/\tilde{g}^2\right)={\rm G}_{1,0}^{0,1}\left[\tilde{g}^2/A\big |^1_{-}\right]$. Then we can apply \cite[Eq. (21)]{adamchik1990algorithm} to calculate the infinite integral of product of Meijer's G-functions involved in $\E_{\tilde{g}}\!\!\left[\exp\left(-A/\tilde{g}^2\right)\right]=\int_{0}^{\infty}\exp\left(-A/\tilde{g}^2\right)f_{\tilde{g}}(\tilde{g})d\tilde{g}$ as \eqref{E_g} shown at the top of the next page. For the ease of notation, hereafter, we denote $\E_{\tilde{g}}\!\!\left[\exp\left(-A/\tilde{g}^2\right)\right]$ by $\mathcal{G}(A)$ for any constant $A$.


Similarly, the second probability term in \eqref{p_out_1} can be obtained, first, by writing $\Pr(\gamma^{(2)}_{\pi_2}>\gamma^{(2)}_{\rm th},\pi_2)=P(\pi_2)-\Pr(\gamma^{(2)}_{\pi_2}<\gamma^{(2)}_{\rm th},\pi_2)$. Then using the symmetry of the problem, it can be shown that $\Pr(\gamma^{(2)}_{\pi_2}<\gamma^{(2)}_{\rm th},\pi_2)$ for $\gamma^{(2)}_{\rm th}<1$ and $\gamma^{(2)}_{\rm th}\geq 1$ can be obtained as \eqref{E2,1} and \eqref{E2,2}, respectively, shown at the top of the next page, where $\mathcal{G}(\cdot)$ is defined in \eqref{E_g}, and $J^{(2)}_{\rm th}=10^{-s/10}\times\gamma^{(2)}_{\rm th}/(1-\gamma^{(2)}_{\rm th})>0$ is defined for $\gamma^{(2)}_{\rm th}<1$. Also, $J^{(2)}_{{\rm th},1}=J^{(2)}_{\rm th}({1+10^{s/10}})$ and $J^{(2)}_{{\rm th},2}=\gamma^{(2)}_{\rm th}+J^{(2)}_{\rm th}({1+\gamma^{(2)}_{\rm th}\times10^{s/10}})$.

\begin{figure*}[!t]
	\normalsize
	\vspace{-0.25cm}
	\!\begin{align}\label{P_1_3}
	\!\!\!\Pr(\gamma^{(1)}_{\pi_2}\!>\!\gamma^{(1)}_{\rm th},\pi_2)\!=\!\frac{1}{1\!+\!10^{s/10}}\!\times\!\exp\!\bigg(\!\!\frac{-\gamma^{(1)}_{\rm th}\sigma^2_{\mathcal{R}}(1\!+\!10^{s/10})}{a_1L_1P}\bigg)\mathcal{G}\!\bigg(\frac{\gamma^{(1)}_{\rm th}C_{\mathcal{D}}(1\!+\!10^{s/10})}{a_1L_1P}\bigg)\prod_{k=1}^{K}\!\frac{a_1L_1P}{a_1L_1P+{\gamma^{(1)}_{\rm th}}L'_kp'_k(1\!+\!10^{s/10})}.
	\end{align}
	\vspace{-0.25cm}
	\hrulefill
\end{figure*}

Moreover, the last probability term in \eqref{p_out_1} can be calculated by first writing
\begin{align}\label{Joint_3}
&\Pr(\gamma^{(1)}_{\pi_2}>\gamma^{(1)}_{\rm th},\pi_2)=\Pr\Big(\gamma^{(1)}_{\rm th}\big[{\boldsymbol{\mathcal{I}}_1}+C_{\mathcal{D}}/(a_1L_1P\tilde{g}^2)\big]\nonumber\\
&\hspace{3cm}<|\tilde{h}_1|^2<|\tilde{h}_2|^2\times10^{-s/10}\Big).
\end{align}
Then using a similar approach to \eqref{Joint1,1}, the closed-form expression for all values of $\gamma^{(1)}_{\rm th}$ can be expressed as \eqref{P_1_3} shown at the top of this page. This completes the closed-form characterization of the outage probability of the first NOMA user $\mathcal{U}_1$.

Finally, the outage probability of the second NOMA user $\mathcal{U}_2$ can be characterized as
\begin{align}\label{p_out_2}
\!\!P_{\rm out}^{(2)}&= 1-\Big[\Pr(\gamma^{(2)}_{\pi_2}>\gamma^{(2)}_{\rm th},\pi_2)+\nonumber\\
&\hspace{-0.5cm}\Pr(\gamma^{(1)}_{\pi_1}>\gamma^{(1)}_{\rm th},\pi_1)\times\Pr(\gamma^{(2)}_{\pi_1}>\gamma^{(2)}_{\rm th},\pi_1)/P(\pi_1)\Big],
\end{align}
where $\Pr(\gamma^{(2)}_{\pi_2}>\gamma^{(2)}_{\rm th},\pi_2)$ and $\Pr(\gamma^{(1)}_{\pi_1}>\gamma^{(1)}_{\rm th},\pi_1)$ have already been calculated, and $\Pr(\gamma^{(2)}_{\pi_1}>\gamma^{(2)}_{\rm th},\pi_1)$ can be obtained as
\begin{align}\label{P_out_2,2}
&\Pr(\gamma^{(2)}_{\pi_1}>\gamma^{(2)}_{\rm th},\pi_1)=(1+10^{-s/10})^{-1}\times\nonumber\\
&\exp\!\bigg(\frac{-\gamma^{(2)}_{\rm th}\sigma^2_{\mathcal{R}}(1\!+\!10^{-s/10})}{a_2L_2P}\bigg)\mathcal{G}\!\bigg(\frac{\gamma^{(2)}_{\rm th}C_{\mathcal{D}}(1\!+\!10^{-s/10})}{a_2L_2P}\bigg)\nonumber\\
&\hspace{1.4cm}\times\prod_{k=1}^{K}\!\frac{a_2L_2P}{a_2L_2P+{\gamma^{(2)}_{\rm th}}L'_kp'_k(1+10^{-s/10})}.
\end{align}
This means that the outage probability of the second NOMA user can be characterized using the preceding analysis by substituting $-s$ for $s$ and appropriate change of indexing $1 \leftrightarrow 2$. This is because the only difference between $\mathcal{U}_1$ and $\mathcal{U}_2$ is that the user with a lower average gain is labeled as the second user, i.e., $a_2L_2=a_1L_1\times10^{-s/10}$.


\subsection{Sum-Rate Outage Analysis}
Assuming that the data rate of the $i$-th NOMA user for the $j$-th decoding order is related to the corresponding SINR as $R^{(i)}_{\pi_j}=\log_2(1+\gamma^{(i)}_{\pi_j})$, $i,j\in\{1,2\}$, then it is easy to verify that the sum rate of the NOMA users, regardless of their decoding order, can be expressed as
\begin{align}\label{R_sum}
\!\!\!\!R_{\Sigma}\!=\!\log_2\!\left(\!1\!+\!\frac{a_1L_1P\tilde{g}^2|\tilde{h}_1|^2+a_2L_2P\tilde{g}^2|\tilde{h}_2|^2}{\sum_{k=1}^{K}{L'_kp'_k}\tilde{g}^2|\tilde{h}'_k|^2+\tilde{g}^2\sigma^2_{\mathcal{R}}+C_{\mathcal{D}}}\!\right)\!.\!
\end{align}
Denoting the fractional term of the logarithm argument in \eqref{R_sum} by $\gamma_{\Sigma}$, the sum-rate outage probability defined as $P_{\rm out}^{\Sigma}=\Pr(\gamma_{\Sigma}<\gamma^{\Sigma}_{\rm th})$, where $\gamma^{\Sigma}_{\rm th}=2^{R^{\Sigma}_{\rm th}}-1$ is the threshold equivalent SINR to achieve the desired sum-rate of $R^{\Sigma}_{\rm th}$, can be expressed as
\begin{align}\label{P_out_sum}
&P_{\rm out}^{\Sigma}=\Pr\Big(|\tilde{h}_1|^2<\gamma^{\Sigma}_{\rm th}\left[\boldsymbol{\mathcal{I}}_1+C_{\mathcal{D}}/(a_1L_1P\tilde{g}^2)\right]\nonumber\\
&\hspace{3.2cm}-|\tilde{h}_2|^2\times10^{-s/10}\Big).
\end{align}

Let $\mathcal{B}$ represent the event $\{|\tilde{h}_2|^2<\gamma^{\Sigma}_{\rm th}\!\times\!10^{s/10}[\boldsymbol{\mathcal{I}}_1\!+\!C_{\mathcal{D}}\!/\!(a_1\!L_1\!P\tilde{g}^2)]\}$, and
\begin{align}
\mathcal{SO}=\{|\tilde{h}_1|^2\!<\!\gamma^{\Sigma}_{\rm th}\!\left[\boldsymbol{\mathcal{I}}_1\!\!+\!C_{\mathcal{D}}\!/\!(a_1\!L_1\!P\tilde{g}^2)\right]\!\!-\!|\tilde{h}_2|^2\!\times\!10^{-s/10}\},
\end{align}
i.e., the sum-rate outage event defined in \eqref{P_out_sum}. Clearly, $\Pr(\mathcal{SO},\mathcal{B}^c)=0$ where $\mathcal{B}^c$ is the complementary event of $\mathcal{B}$. Therefore, using the law of total probability, $P_{\rm out}^{\Sigma}$ can be expressed as $P_{\rm out}^{\Sigma}=\Pr(\mathcal{SO},\mathcal{B})$ which is calculated in a closed-form as \eqref{Pout_sum} at the top of the next page. 

\begin{figure*}[!t]
	\normalsize
	\vspace{-0.25cm}
	\begin{align}\label{Pout_sum}
	P_{\rm out}^{\Sigma}=&\E_{|\tilde{h}_2|^2<\gamma^{\Sigma}_{\rm th}\!\times\!{10^{s/10}}\big[\boldsymbol{\mathcal{I}}_1+C_{\mathcal{D}}/(a_1L_1P\tilde{g}^2)\big]}\left[1-\exp\left(-\gamma^{\Sigma}_{\rm th}\!\left[\boldsymbol{\mathcal{I}}_1+C_{\mathcal{D}}/(a_1L_1P\tilde{g}^2)\right]+|\tilde{h}_2|^2\times10^{-s/10}\right)\right]\nonumber\\
	=&1+\frac{1}{10^{s/10}-1}\times\exp\!\bigg(\frac{-\sigma^2_{\mathcal{R}}\gamma^{\Sigma}_{\rm th}\times10^{s/10}}{a_1L_1P}\bigg){\mathcal{G}}\!\bigg(\frac{C_{\mathcal{D}}\gamma^{\Sigma}_{\rm th}\times10^{s/10}}{a_1L_1P}\bigg)\prod_{k=1}^{K}\frac{a_1L_1P}{a_1L_1P+L'_kp'_k\gamma^{\Sigma}_{\rm th}\times10^{s/10}}\nonumber\\
	&{\hspace{2cm}}-\frac{10^{s/10}}{10^{s/10}-1}\times\exp\!\bigg(\frac{-\gamma^{\Sigma}_{\rm th}\sigma^2_{\mathcal{R}}}{a_1L_1P}\bigg){\mathcal{G}}\!\bigg(\frac{\gamma^{\Sigma}_{\rm th}C_{\mathcal{D}}}{a_1L_1P}\bigg)\prod_{k=1}^{K}\frac{a_1L_1P}{a_1L_1P+\gamma^{\Sigma}_{\rm th}L'_kp'_k}.
	\end{align}
	\hrulefill
	\vspace{-0.25cm}
\end{figure*}

It is worth remarking at this point that in the special case of absence of multiuser interference (except the NOMA users themselves), one can obtain the outage probability closed-form expressions by substituting $L'_kp'_k=0$, $\forall k=1,2,...,K$, which replaces all the product terms of the form $\prod_{k=1}^K[\cdot]$ by $1$.
\section{Numerical Results}
In this section, we present the numerical results to evaluate the performance of uplink NOMA over mixed RF-FSO systems, and corroborate the correctness of the derived outage probability closed forms. Some of the  parameters considered for simulations are listed in Table I. For the multiuser interference, we consider the product of $L'_kp'_k$, $k=1,2,...,K$, to be the $k$-th element of the vector $K_{\mathcal{I}}P_0L_2{\boldsymbol{u}_{10}}$ where $P_0=1$ \si{mW}, $K_{\mathcal{I}}\geq0$ is a constant to define the upper bound of the received power from each interfering user as a factor of $P_0L_2$, and ${\boldsymbol{u}_{10}}=(0.6957,0.6279,0.4504,0.4736,0.9497,0.0835,0.2798,0.4470,\\0.5876,0.8776)$ is a length-10 vector of uniformly generated numbers over the interval $(0,1)$.

\begin{table}[t]
\centering
\caption{Some of the important parameters used for simulations.}
\label{T2}
\begin{tabular}{M{2.33in}||M{0.5in}}

	Coefficient & Value\\
	\hline \hline
Responsivity of the photodetector, $\rho$ & $0.5$ \si{V^{-1}}\\\hline
Electrical-to-optical conversion coefficient, $\eta$ & $1$\\\hline
Receiver aperture radius, $r$ & $10$ \si{cm}\\\hline
Transmitter beam divergence angle, $\phi$ & $2$ \si{mrad}\\\hline
Noise power at the relay RF receiver, $\sigma^2_{\mathcal{R}}$ &$-80$ \si{dBm} \\\hline
Noise variance at the destination FSO receiver, $\sigma^2_{\mathcal{D}}$ & $10^{-14}$ \si{A^2}\\\hline
Number of interfering users to relay, $K$ & $10$\\
\hline
Number of iterations for numerical simulations, $N_t$ & $10^{6}$\\\hline
Gamma-Gamma turbulence parameters, $(\alpha,\beta)$ & $(10,5)$\\\hline
Length of the FSO backhaul link, $d_{\mathcal{RD}}$& $800$ \si{m}\\\hline
Weather-dependent attenuation coefficient, $\kappa$& $0.02$ \si{m^{-1}}\\\hline
\end{tabular}
\end{table}

\begin{figure}\label{fig1}
	\centering
	\includegraphics[width=3.6in]{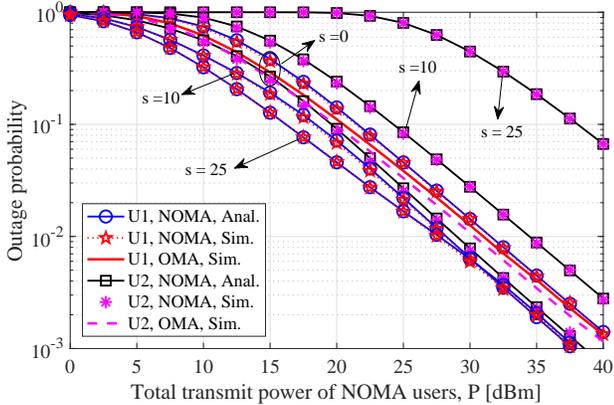}
	\caption{Individual-rate outage probability results of the mixed RF-FSO NOMA system for three values of power back-off step $s=0$, $10$, and $25$ \si{dB}. The other specific parameters are $\gamma_{\rm th}^{(1)}=0.8$, $\gamma_{\rm th}^{(2)}=0.4$, $L_1=2\times10^{-7}$, $L_2=10^{-7}$, $K_{\mathcal{I}}=1$, $\zeta=2$, and $G=100$.}
	\vspace{-0.15in}
\end{figure}
Figure 1 shows the individual-rate outage performance of the uplink mixed RF-FSO NOMA system for three different values of the power back-off step. For $s=0$ we will have $a_1L_1=a_2L_2$; therefore, one should expect a lower outage probability for the second NOMA user given its lower threshold SINR. However, by increasing $s$ a larger fraction of power will be assigned to the first NOMA user, and $\mathcal{U}_1$ achieves lower outage probabilities even if it has a larger SINR threshold. As a consequence, increasing $s$ will decrease the outage probability of $\mathcal{U}_1$ and increase the outage probability of $\mathcal{U}_2$. Moreover, the excellent match between the analytical results and Monte-Carlo numerical simulations corroborate the correctness of the derived closed-form expressions for the individual-rate outage probabilities. 

The comparison between NOMA and OMA is also depicted in Figure 1. To do so, we assume, for OMA operation, that the total transmission time is equally divided between the two users and each user employs the entire transmission power $P$ during its corresponding time slot. Then it is easy to verify that, in order to achieve the target data rate $R_{\rm th}^{(i)}=\log_2(1+\gamma_{\rm th}^{(i)})$, $i\in\{1,2\}$, each $i$-th OMA user has to satisfy the threshold SNR of $\gamma^{(i)}_{{\rm th},\rm OMA}=(1+\gamma_{\rm th}^{(i)})^2-1$. It is observed that NOMA operation is in favor of the first user except for very small values of $s$ while the second user experiences an opposite situation. 

\begin{figure}\label{fig2}
	\centering
	\includegraphics[width=3.6in]{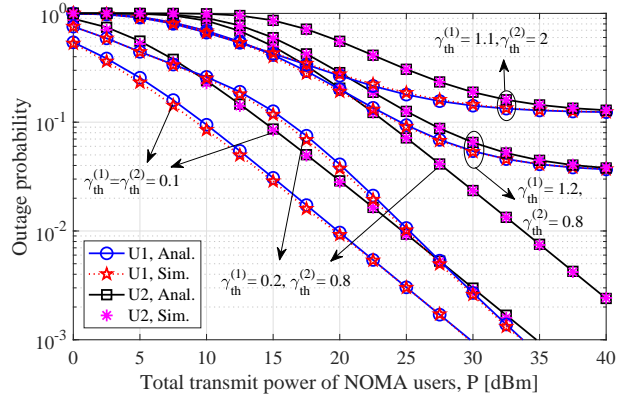}
	\caption{Individual-rate outage probability results of the mixed RF-FSO NOMA system for power back-off step  $s=5$ \si{dB}, and different values of threshold SINRs. The other parameters are  the same as Figure 1.}
	\vspace{-0.15in}
\end{figure}
Figure 2 illustrates the individual-rate outage performance of the system for $s=5$ \si{dB} and different values of threshold SINRs. As expected, outage performance degrades with increasing the threshold SINRs. More importantly, the induced interference between NOMA users due to the non-orthogonal operation limits the outage performance for large values of threshold SINRs and prevents achieving small enough outage probabilities even for large values of the transmitted power. Consequently,  the system performance saturates where the saturation limit is larger for the larger values of threshold SINRs.

\begin{figure}\label{fig3}
	\centering
	\includegraphics[width=3.6in]{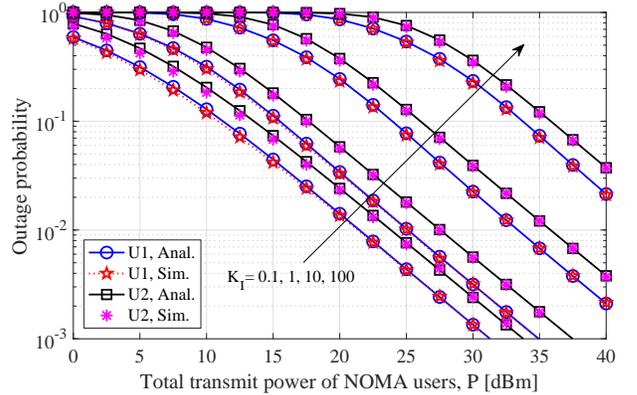}
	\caption{Individual-rate outage probability results of the mixed RF-FSO NOMA system for power back-off step  $s=5$ \si{dB}, $\gamma_{\rm th}^{(1)}=0.7$, $\gamma_{\rm th}^{(2)}=0.4$, $L_1=10^{-6}$, $L_2=2\times10^{-7}$, $\zeta=2$, $G=100$, and different values of $K_{\mathcal{I}}$.}
	\vspace{-0.15in}
\end{figure}
The impact of multiuser interference on the individual-rate outage performance of NOMA users is investigated in Figure 3 where a significant performance degradation is observed for relatively strong interference regimes. Note that even for $K_{\mathcal{I}}=1$ which is equivalent to having interfering users with the same path loss gain of $L_2$ as the second NOMA user and transmit powers bounded by $P_0=0$ \si{dBm}, one can observe  about $4$ \si{dB} of performance degradation compared to $K_{\mathcal{I}}=0.1$ at the outage probability $10^{-3}$.

\begin{figure}\label{fig4}
	\centering
	\includegraphics[width=3.6in]{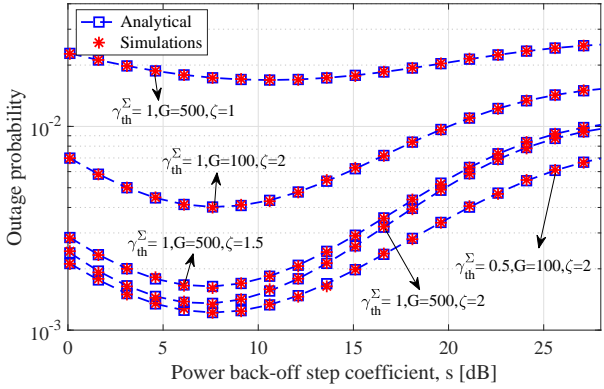}
	\caption{Sum-rate outage probability results of the mixed RF-FSO NOMA system for  $L_1=10^{-6}$, $L_2=2\times10^{-7}$, $P=20$ \si{dBm}, $K_{\mathcal{I}}=1$, and different values of $\gamma_{\rm th}^{\Sigma}$, $\zeta$, and $G$.}
	\vspace{-0.15in}
\end{figure}
The sum-rate outage performance of the system is characterized in Figure 4 for $P=20$ \si{dBm}  and different values of $\gamma_{\rm th}^{\Sigma}$, $\zeta$, and $G$. As expected, the outage performance increases for larger values of the threshold SINR $\gamma_{\rm th}^{\Sigma}$, smaller values of the relay gain $G$, and higher pointing errors (equivalently, smaller $\zeta$). Furthermore, given any set of system parameters, there is unique power back-off step $s^*$ minimizing the sum-rate outage probability. However, such a $s^*$ is not necessarily the best operation point as such an operation region may depend to the individual outage probabilities and achievable rates and not only to the sum-rate outage probability. More importantly, it is observed that the performance degradation due to the laser beam misalignment can be quite remarkable. This necessitates hybrid design of the backhaul link to incorporate an RF or millimeter wave link as a backup to assist the FSO backhaul link in the case of poor transmission quality of the FSO link. Such designs are necessary to guarantee the users requirements for high reliability and low latency, and will be explored in our future work. 

\section{Conclusions}
In this paper, we derived the closed-form expressions for the individual- and sum-rate outage probabilities of dual-hop uplink NOMA over mixed RF-FSO  systems with Rayleigh fading for the users-relay access links, multiuser interference to the relay, Gamma-Gamma turbulence with pointing error for the FSO backhaul link, and dynamic-order decoding at the destination. This work can be regarded as an initial attempt to incorporate ultra-high-throughput FSO links as an effective backhauling solution to meet the  ever-increasing demand of users for higher data rates and the stringent requirements of reliability and latency for variety of emerging applications. The analysis in this paper are performed for general cases and their validity is verified through extensive numerical results. 


\begin{thebibliography}{10}
	\providecommand{\url}[1]{#1}
	\csname url@samestyle\endcsname
	\providecommand{\newblock}{\relax}
	\providecommand{\bibinfo}[2]{#2}
	\providecommand{\BIBentrySTDinterwordspacing}{\spaceskip=0pt\relax}
	\providecommand{\BIBentryALTinterwordstretchfactor}{4}
	\providecommand{\BIBentryALTinterwordspacing}{\spaceskip=\fontdimen2\font plus
		\BIBentryALTinterwordstretchfactor\fontdimen3\font minus
		\fontdimen4\font\relax}
	\providecommand{\BIBforeignlanguage}[2]{{%
			\expandafter\ifx\csname l@#1\endcsname\relax
			\typeout{** WARNING: IEEEtran.bst: No hyphenation pattern has been}%
			\typeout{** loaded for the language `#1'. Using the pattern for}%
			\typeout{** the default language instead.}%
			\else
			\language=\csname l@#1\endcsname
			\fi
			#2}}
	\providecommand{\BIBdecl}{\relax}
	\BIBdecl
	
	\bibitem{dai2015non}
	L.~Dai, B.~Wang, Y.~Yuan, S.~Han, I.~Chih-Lin, and Z.~Wang, ``Non-orthogonal
	multiple access for {5G}: solutions, challenges, opportunities, and future
	research trends,'' \emph{IEEE Commun. Mag.}, vol.~53, no.~9, pp. 74--81,
	2015.
	
	\bibitem{ding2017survey}
	Z.~Ding, X.~Lei, G.~K. Karagiannidis, R.~Schober, J.~Yuan, and V.~K. Bhargava,
	``A survey on non-orthogonal multiple access for {5G} networks: Research
	challenges and future trends,'' \emph{IEEE J. Sel. Areas Commun.}, vol.~35,
	no.~10, pp. 2181--2195, 2017.
	
	\bibitem{kim2015capacity}
	J.-B. Kim and I.-H. Lee, ``Capacity analysis of cooperative relaying systems
	using non-orthogonal multiple access,'' \emph{IEEE Commun. Lett.}, vol.~19,
	no.~11, pp. 1949--1952, 2015.
	
	\bibitem{zhong2016non}
	C.~Zhong and Z.~Zhang, ``Non-orthogonal multiple access with cooperative
	full-duplex relaying,'' \emph{IEEE Commun. Lett.}, vol.~20, no.~12, pp.
	2478--2481, 2016.
	
	\bibitem{men2015non}
	J.~Men and J.~Ge, ``Non-orthogonal multiple access for multiple-antenna
	relaying networks,'' \emph{IEEE Commun. Lett.}, vol.~19, no.~10, pp.
	1686--1689, 2015.
	
	\bibitem{men2017performance}
	J.~Men, J.~Ge, and C.~Zhang, ``Performance analysis of nonorthogonal multiple
	access for relaying networks over {Nakagami-$m$} fading channels,''
	\emph{IEEE Trans. Veh. Technol.}, vol.~66, no.~2, pp. 1200--1208, 2017.
	
	\bibitem{kim2015coordinated}
	J.-B. Kim and I.-H. Lee, ``Non-orthogonal multiple access in coordinated direct
	and relay transmission,'' \emph{IEEE Commun. Lett.}, vol.~19, no.~11, pp.
	2037--2040, 2015.
	
	\bibitem{kader2018coordinated}
	M.~F. Kader and S.~Y. Shin, ``Coordinated direct and relay transmission using
	uplink {NOMA},'' \emph{IEEE Wireless Commun. Lett.}, vol.~7, no.~3, pp.
	400--403, 2018.
	
	\bibitem{liu2016hybrid}
	Y.~Liu, G.~Pan, H.~Zhang, and M.~Song, ``Hybrid decode-forward \&
	amplify-forward relaying with non-orthogonal multiple access,'' \emph{IEEE
		Access}, vol.~4, pp. 4912--4921, 2016.
	
	\bibitem{xiao2018forwarding}
	Y.~Xiao, L.~Hao, Z.~Ma, Z.~Ding, Z.~Zhang, and P.~Fan, ``Forwarding strategy
	selection in dual-hop {NOMA} relaying systems,'' \emph{IEEE Commun. Lett., to
		appear}, Feb. 2018.
	
	\bibitem{jamali2018low}
	M.~V. Jamali and H.~Mahdavifar, ``A low-complexity recursive approach toward
	code-domain {NOMA} for massive communications,'' \emph{arXiv preprint
		arXiv:1804.05242}, 2018.
	
	\bibitem{ge20145g}
	X.~Ge, H.~Cheng, M.~Guizani, and T.~Han, ``{5G} wireless backhaul networks:
	challenges and research advances,'' \emph{IEEE Network}, vol.~28, no.~6, pp.
	6--11, 2014.
	
	\bibitem{demers2011survey}
	F.~Demers, H.~Yanikomeroglu, and M.~St-Hilaire, ``A survey of opportunities for
	free space optics in next generation cellular networks,'' in
	\emph{Communication Networks and Services Research Conference (CNSR), 2011
		Ninth Annual}.\hskip 1em plus 0.5em minus 0.4em\relax IEEE, 2011, pp.
	210--216.
	
	\bibitem{andrews2017modeling}
	J.~G. Andrews, T.~Bai, M.~N. Kulkarni, A.~Alkhateeb, A.~K. Gupta, and R.~W.
	Heath, ``Modeling and analyzing millimeter wave cellular systems,''
	\emph{IEEE Trans. Commun.}, vol.~65, no.~1, pp. 403--430, 2017.
	
	\bibitem{khalighi2014survey}
	M.~A. Khalighi and M.~Uysal, ``Survey on free space optical communication: A
	communication theory perspective,'' \emph{IEEE Commun. Surveys Tuts},
	vol.~16, no.~4, pp. 2231--2258, 2014.
	
	\bibitem{gao2017theoretical}
	Y.~Gao, B.~Xia, K.~Xiao, Z.~Chen, X.~Li, and S.~Zhang, ``Theoretical analysis
	of the dynamic decode ordering {SIC} receiver for uplink {NOMA} systems,''
	\emph{IEEE Commun. Lett.}, vol.~21, no.~10, pp. 2246--2249, 2017.
	
	\bibitem{najafi2018non}
	M.~Najafi, V.~Jamali, P.~D. Diamantoulakis, G.~K. Karagiannidis, and
	R.~Schober, ``Non-orthogonal multiple access for {FSO} backhauling,'' in
	\emph{Wireless Communications and Networking Conference (WCNC), 2018
		IEEE}.\hskip 1em plus 0.5em minus 0.4em\relax IEEE, 2018, pp. 1--6.
	
	\bibitem{jamali2016link}
	V.~Jamali, D.~S. Michalopoulos, M.~Uysal, and R.~Schober, ``Link allocation for
	multiuser systems with hybrid {RF/FSO} backhaul: Delay-limited and
	delay-tolerant designs,'' \emph{IEEE Trans. Wireless Commun.}, vol.~15,
	no.~5, pp. 3281--3295, 2016.
	
	\bibitem{lee2011performance}
	E.~Lee, J.~Park, D.~Han, and G.~Yoon, ``Performance analysis of the asymmetric
	dual-hop relay transmission with mixed {RF/FSO} links,'' \emph{IEEE Photonics
		Technol. Lett.}, vol.~23, no.~21, pp. 1642--1644, 2011.
	
	\bibitem{sandalidis2009optical}
	H.~G. Sandalidis, T.~A. Tsiftsis, and G.~K. Karagiannidis, ``Optical wireless
	communications with heterodyne detection over turbulence channels with
	pointing errors,'' \emph{J. Lightw. Technol.}, vol.~27, no.~20, pp.
	4440--4445, 2009.
	
	\bibitem{gradshteyn2014table}
	I.~S. Gradshteyn and I.~M. Ryzhik, \emph{Table of integrals, series, and
		products}.\hskip 1em plus 0.5em minus 0.4em\relax Academic Press, 2007.
	
	\bibitem{zhang2016uplink}
	N.~Zhang, J.~Wang, G.~Kang, and Y.~Liu, ``Uplink nonorthogonal multiple access
	in {5G} systems,'' \emph{IEEE Commun. Lett.}, vol.~20, no.~3, pp. 458--461,
	2016.
	
	\bibitem{yang2016general}
	Z.~Yang, Z.~Ding, P.~Fan, and N.~Al-Dhahir, ``A general power allocation scheme
	to guarantee quality of service in downlink and uplink {NOMA} systems,''
	\emph{IEEE Trans. Wireless Commun.}, vol.~15, no.~11, pp. 7244--7257, 2016.
	
	\bibitem{adamchik1990algorithm}
	V.~Adamchik and O.~Marichev, ``The algorithm for calculating integrals of
	hypergeometric type functions and its realization in reduce system,'' in
	\emph{Proc. Int. Conf. Symbolic and Algebraic Computation}.\hskip 1em plus
	0.5em minus 0.4em\relax ACM, Tokyo, Japan, 1990, pp. 212--224.
	
\end{thebibliography}
\end{document}